\documentclass{article} 
\usepackage{iclr2020_conference,times}

\usepackage{hyperref}
\usepackage{url}

\iclrfinalcopy

\usepackage[ruled,linesnumbered,vlined]{algorithm2e}
\usepackage{graphicx}
\usepackage{caption}
\usepackage{subcaption}
\usepackage{hyperref}
\usepackage{amsmath}
\usepackage[utf8]{inputenc}
\usepackage{amsfonts}
\usepackage{hyperref}
\usepackage{siunitx}
\usepackage{xcolor}
\usepackage{amssymb}

\newcommand{\sio}{SiO\textsubscript{2}}
\newcommand{\tio}{TiO\textsubscript{2}}
\newcommand{\mgf}{MgF\textsubscript{2}}
\newcommand{\alo}{Al\textsubscript{2}O\textsubscript{3}}

\title{Automated Optical Mutli-layer Design via \\Deep Reinforcement Learning}

\begin{document}
	
\author{Haozhu Wang\thanks{Corresponding author: \texttt{hzwang@umich.edu}}, Zeyu Zheng \& L. Jay Guo  \\
	EECS Department, University of Michigan\\
	Ann Arbor, MI 48105, USA \\
	\texttt{\{hzwang, zeyu, guo\}@umich.edu} \\
	\And
	Chengang Ji \\
	Inlight Technology Co., Ltd.,\\
	Ningbo, Zhejiang, China\\
	\texttt{jichg@umich.edu} \\ 
}

\maketitle

\begin{abstract}
Optical multi-layer thin films are widely used in optical and energy applications requiring photonic designs. Engineers often design such structures based on their physical intuition. However, solely relying on human experts can be time-consuming and may lead to sub-optimal designs, especially when the design space is large. In this work, we frame the multi-layer optical design task as a sequence generation problem. A deep sequence generation network is proposed for efficiently generating optical layer sequences. We train the deep sequence generation network with proximal policy optimization to generate multi-layer structures with desired properties. The proposed method is applied to two energy applications. Our algorithm successfully discovered high-performance designs, outperforming structures designed by human experts in task 1, and a state-of-the-art memetic algorithm in task 2.
\end{abstract}

Optical multi-layer films have been widely used in many applications, such as broadband filtering \cite{yang2016compact}, photovoltaics \cite{agrawal2008broadband}, radiative cooling \cite{raman2014passive}, and structural colors \cite{li2018photonic}. The design of optical multi-layer films is a combinatorial optimization problem that requires one to choose the best combination of materials and layer thicknesses to form a multi-layer structure. Researchers and engineers often make such designs based on their physical intuition. However, a completely human-based design process is slow and often leads to sub-optimal designs, especially when the design space is enormous. Thus, computational methods for designing optical multi-layer structures, including evolutionary algorithms \cite{schubert2008design, shi2017optimization}, needle optimization \cite{tikhonravov1996application}, and particle swarm optimization \cite{rabady2014global}, have been proposed to tackle this problem. All of these previous methods frame the optical design task as an optimization problem and aim to synthesize a structure that meets user-specified design criteria. However, these methods for optical design are based entirely on heuristic search, i.e., they do not learn a model to solve the design problems. When the heuristic approach is sub-optimal for a task, the search process may fail to identify a high-performance design. 

\begin{figure}[ht]
    \centering
    \includegraphics[width=0.7\textwidth]{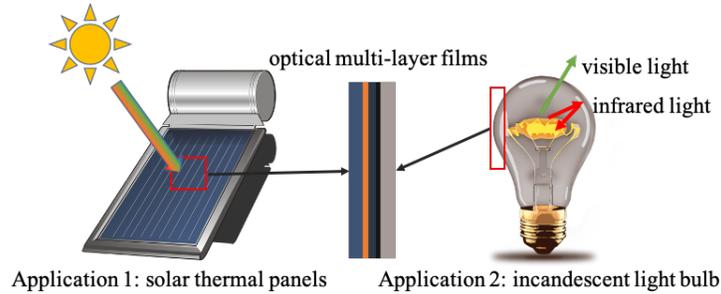}
    \caption{Two energy applications of optical multi-layer films. For solar thermal panels, we can use multi-layer films as ultra-wideband absorbers to enhance light absorption efficiency. For incandescent light bulbs, we can coat multi-layer films on them to improve luminous efficiency by reflecting infrared light while transmitting visible light.}
    \label{fig:application_illustration}
\end{figure}

In contrast, deep reinforcement learning (DRL) is a learning framework that learns to solve complex tasks through an trial-and-error process. It is proven to be highly scalable for solving large-scale and complicated tasks \cite{silver2017mastering, vinyals2019grandmaster}. Researchers have successfully applied DRL to various combinatorial optimization problems \cite{bello2016neural, khalil2017learning, mirhoseini2017device, mirhoseini2020chip}. Unlike heuristic-based search, reinforcement learning methods learn a model using the reward signal \cite{sutton2018reinforcement} and do not depend on hand-crafted heuristics. On some combinatorial optimization tasks, DRL has been shown to outperform classic heuristic search methods \cite{Lu2020A}. Recently, researchers applied DRL on designing optical devices with a structure template \cite{sajedian2019optimisation, sajedian2019double}, where the number of layers is fixed. However, when designing the optical multi-layer films, we often do not know the optimal structure template. Thus, the previous DRL approaches are not suitable for multi-layer designs. In addition to DRL, deep learning-enabled inverse design methods have seen great development in recent years \cite{ma2018deep, liu2018training, liu2018generative}. These inverse design models learn a mapping between design targets and design parameters using a static training set, which allows users to efficiently retrieve designs that match design targets. However, if a design target does not lie within the training datasets used for training the inverse design model, we will not be able to obtain the corresponding design using the inverse design model. For our performance optimization task, the optimal design is often not covered by a static training dataset. Otherwise, it would mean that the optimization task has already been solved through the training dataset collection process. Thus, reinforcement learning is more suitable than deep-learning-based inverse design methods when users want to optimize the design performance.

Because the multi-layer optical design task is equivalent to a sequence generation problem, we propose a DRL method called Optical Multi-layer Proximal Policy Optimization (OML-PPO) that can generate near-optimal multi-layer structures. The proposed method uses a state-of-the-art DRL algorithm PPO to train a deep recurrent neural network that outputs near-optimal optical designs. We introduce two novel designs for the deep recurrent neural network to allow it to efficiently explore the design space. With an ablation study, we show that the proposed neural network architecture enables the RL agent to explore the design space efficiently. 

We applied the proposed method to two optical design tasks that are relevant to energy applications (Figure \ref{fig:application_illustration}): 1) ultra-wideband absorbers that can enhance light-harvesting efficiency, e.g. for thermal photovoltaics and photothermal energy conversion 2) incandescent light bulb filters that can improve light bulb efficiency in emitting visible light. On the task of designing ultra-wideband absorbers, we show that OML-PPO can reliably discover high-performance designs. A 5-layer structure with 97.64\% average absorption over the wavelength range [400, 2000] nm is discovered by OML-PPO, outperforming a previously reported structure using the same number of layers with 95.37\% average absorption. When applied to generate absorbers with more layers, OML-PPO discovers a 14-layer structure that achieves near-perfect 99.24\% average absorption. We also applied our method to design a 42-layer incandescent light bulb filter and achieved an enhancement factor of 16.60, which is 8.5\% higher than a 41-layer structure designed by a state-of-the-art memetic algorithm. Our results demonstrate that the proposed algorithm is efficient at discovering near-optimal designs and is scalable to complicated design problems. We summarize our contributions: 
\begin{enumerate}
    \item We frame the multi-layer optical design task as a sequence generation problem and develop a DRL method (OML-PPO) for solving this task.
    \item We propose a novel deep sequence generation network that allows efficient exploration of the optical design space.
    \item On two optical design tasks, we demonstrate that our method is effective in discovering near-optimal solutions for complicated design tasks.
\end{enumerate}

\section{Related Work}

Researchers have developed reinforcement learning methods for solving various combinatorial optimization problems. In \cite{bello2016neural}, the authors trained a Pointer Network \cite{vinyals2015pointer} to solve the Traveling Salesman Problem (TSP). Khalil et al. \cite{khalil2017learning} combined graph embedding and RL for solving a diverse set of combinatorial optimization problems including the Minimum Vertex Cover, Maximum Cut, and TSP. Chen and Tian \cite{chen2019learning} proposed a method to learn policies that can rewrite the heuristics in existing solvers for combinatorial optimization problems. Lu et al. \cite{Lu2020A} showed that RL-based method could outperform a classic operation research algorithm in terms of both average cost and time efficiency.

Many real-life applications can be formalized as sequence generation problems \cite{li2016deep, popova2018deep, Angermueller2020Model-based, mirhoseini2020chip}. In \cite{li2016deep}, the authors integrated RL and seq2seq to automatically generate a response by simulating the dialogue between two agents. In \cite{Angermueller2020Model-based}, the authors proposed a model-based variant of PPO to deal with the large-batch, low round setting for biological sequence design \cite{Angermueller2020Model-based}. Mirhoseini et al. \cite{mirhoseini2020chip} combined graph neural networks with RL for sequentially placing devices on a chip. These previous works all trained sequence generation models using policy gradient algorithms. In this work, we introduced a sequence generation network architecture tailored to the optical design task. Additionally, we combined local search with DRL for finetuning the thicknesses of the generated layers.

Deep-learning-based inverse design \cite{ma2018deep, liu2018training, liu2018generative} has been gaining popularity in recent years. In \cite{ma2018deep}, the authors trained convolutional neural networks to directly predict design parameters using the design target as the input to the network. Liu et al. \cite{liu2018generative} trained a generative adversarial network (GAN) to inversely design optical devices by generating 2D shapes of the optical structure. However, these approaches all rely on a curated training set that contains diverse examples. When our goal is to push the performance limit of certain devices, the near-optimal structures is unlikely to be within the training data distribution. Thus, these static methods are not appropriate for optimizing design performances. Our proposed method tackles this problem by actively searching the design space to generate high-performance designs via reinforcement learning. In \cite{jiang2019free}, the authors also developed an active search process by adding additional high-quality data to augment the initial training set. However, their approach requires the users to retrain the neural network with the augmented dataset while our RL-based method accomplishes the design task within one training process.

\section{Methods}

Multi-layer films can be treated as sequences. Each layer is represented as $s_l=(m_l, d_l)$. We can represent such a structure with $N$ layers as $\mathcal{S} = \{(m_0, d_0), (m_1, d_1), (m_2, d_2), \cdots, (m_{N-1}, d_{N-1})\}$, where $m_l$ and $d_l$ denote the material and the thickness of the $l$-th layer (counting from the top), respectively. When designing optical multi-layer films, we hope to synthesize a sequence that has the desired target spectral response $\tilde{T}$ . Thus, the design task is equivalent to a sequence generation problem, where we generate $m$ and $d$ in each step. Generation tasks such as dialogue generation \cite{li2016deep}, molecule generation \cite{popova2018deep}, and biological sequence generation \cite{Angermueller2020Model-based} have been widely studied by machine learning researchers. In these works, researchers train a neural network as a generator for synthesizing sequences. Because we do not have ground-truth data for optimal design tasks, we apply reinforcement learning \cite{sutton2018reinforcement} to train the sequence generator. 

\subsection{Sequence generation network}
To generate the optical layer sequences, we use a recurrent neural network (RNN) \cite{hochreiter1997long}. Unlike simple feed-forward neural networks, RNNs maintain a hidden state $h$ that contains useful information from the history of the sequence. Thus, RNNs are suitable for tasks that require memorizing history and have been widely used in sequence generation tasks \cite{graves2013generating}. Gated recurrent units (GRUs) \cite{chung2014empirical} and long short-term memory networks (LSTMs) \cite{hochreiter1997long} are two popular variants of RNNs. Researchers have previously found that the empirical performance of GRUs and LSTMs is similar. Because GRUs have a simpler structure than LSTMs and require fewer parameters to train, we choose to use a GRU for generating the optical multi-layer structures. Similar to sampling words from a dictionary when generating a sentence, we sample the material $m_l$ from a fixed set of materials $\mathcal{M}$ for each layer. Though the thickness $d_l$ is intrinsically a continuous variable, we choose to sample the thickness from a set of discrete values $\mathcal{D}$ to reduce the size of the exploration space. Later, we apply quasi-Newton methods \cite{zhu1997algorithm} to finetune the layer thicknesses of the generated structure for further performance improvement. 

\begin{figure}[ht]
    \centering
    \includegraphics[height=1.1in]{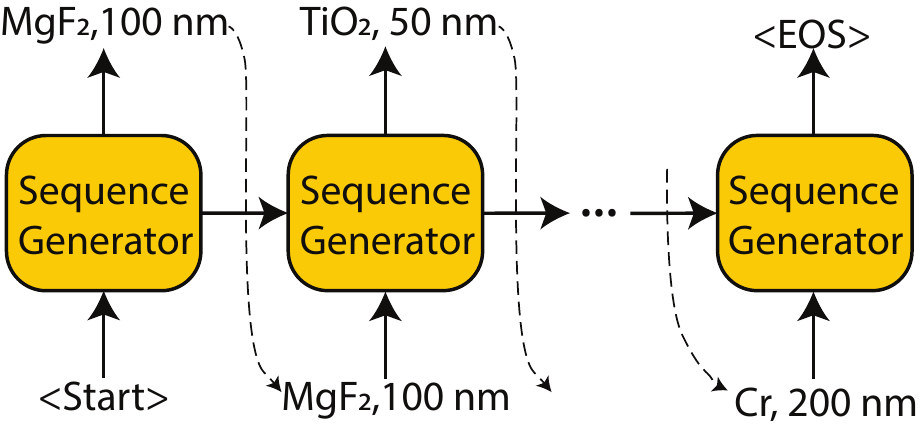}
    \caption{Optical multi-layer design as sequence generation. The generation process will stop when either the EOS token is sampled, or the length of the sequence reaches the maximum allowed length $L$.}
    \label{fig:generation}
\end{figure}

Our optical multi-layer sequence generation network consists of a GRU and two multi-layer perceptrons (MLPs) \cite{goodfellow2016deep}. At generation step $l$, the GRU takes its own output from the previous step $s_{l-1}=(m_{l-1}, d_{l-1})$ and the previous hidden state $h_l$ as the inputs to compute the hidden state $h_l$. This auto-regressive generation process allows the GRU to remember what has been generated so far. To generate the material and thickness for layer $l$, the hidden state $h_l$ of the GRU is inputted to two MLPs. One of the MLPs outputs logits vector $\sigma_{m_l}\in \mathbb{R}^{|\mathcal{M}|+1}$ corresponding to all possible materials and an end-of-sequence token (EOS). The other MLP outputs a thickness logits vector $\sigma_{d_l} \in \mathbb{R}^{|\mathcal{D}|}$ corresponding to all allowable thicknesses in the set $\mathcal{D}$. Then, we transform these logits vectors with the \textit{softmax} function to obtain proper probability distributions. Finally, the material and thickness are sampled from their corresponding distributions. The generation process will stop either when the length reaches the maximum length $L$ set by the user or when the EOS token is sampled. Thus, the number of layers $N$ of a generated structure is always lower than or equal to the maximum sequence length $L$. The process for generating a sequence is illustrated in Figure. \ref{fig:generation}.

\begin{figure}[ht]
    \begin{subfigure}[c]{0.5\textwidth}
        \centering
        \includegraphics[trim={0cm 0 2.0cm 0},clip,height=2.5in]{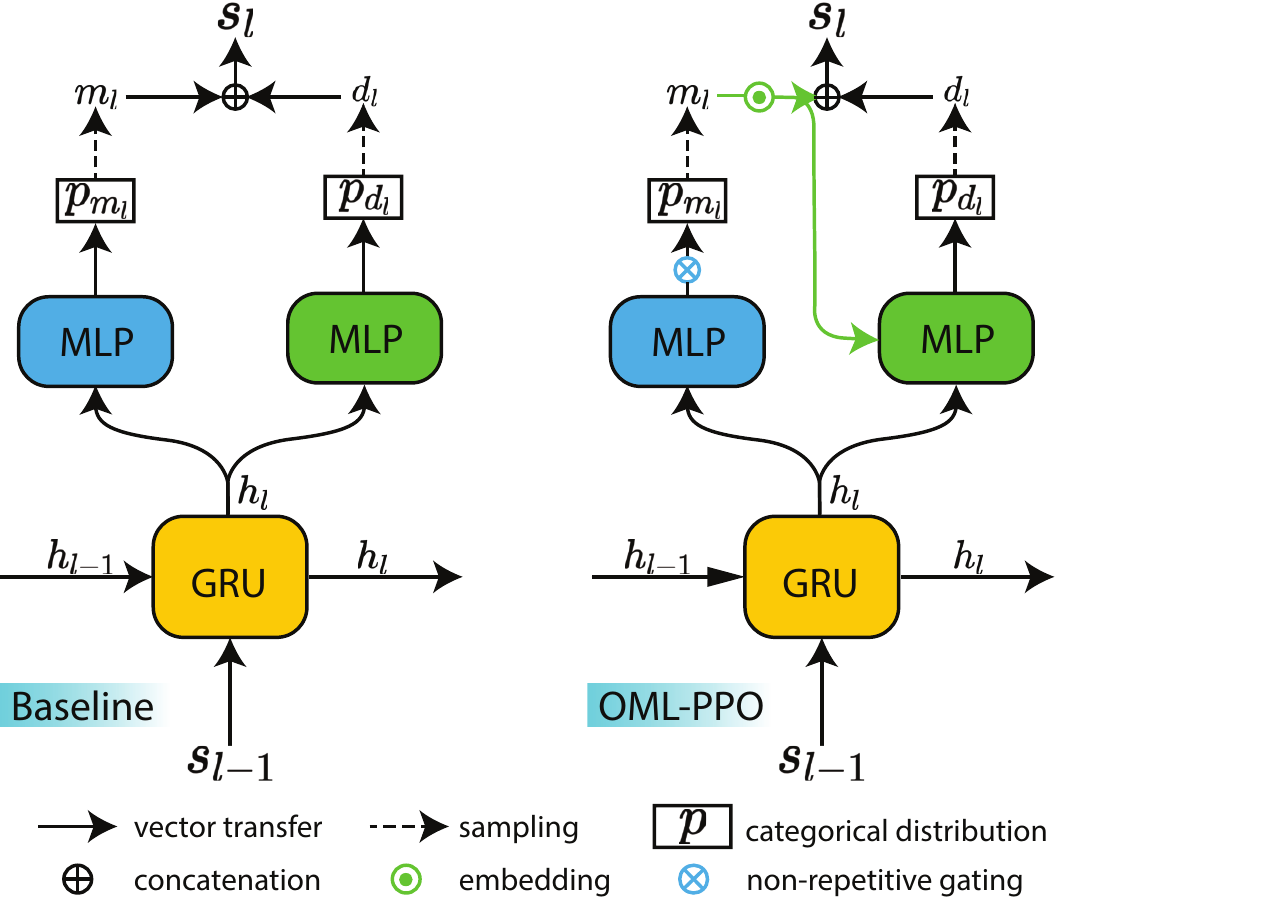}
        \caption{}
        \label{fig:architecure}
    \end{subfigure}
    ~
    \begin{subfigure}[c]{0.4\textwidth}
        \centering
        \includegraphics[height=2in]{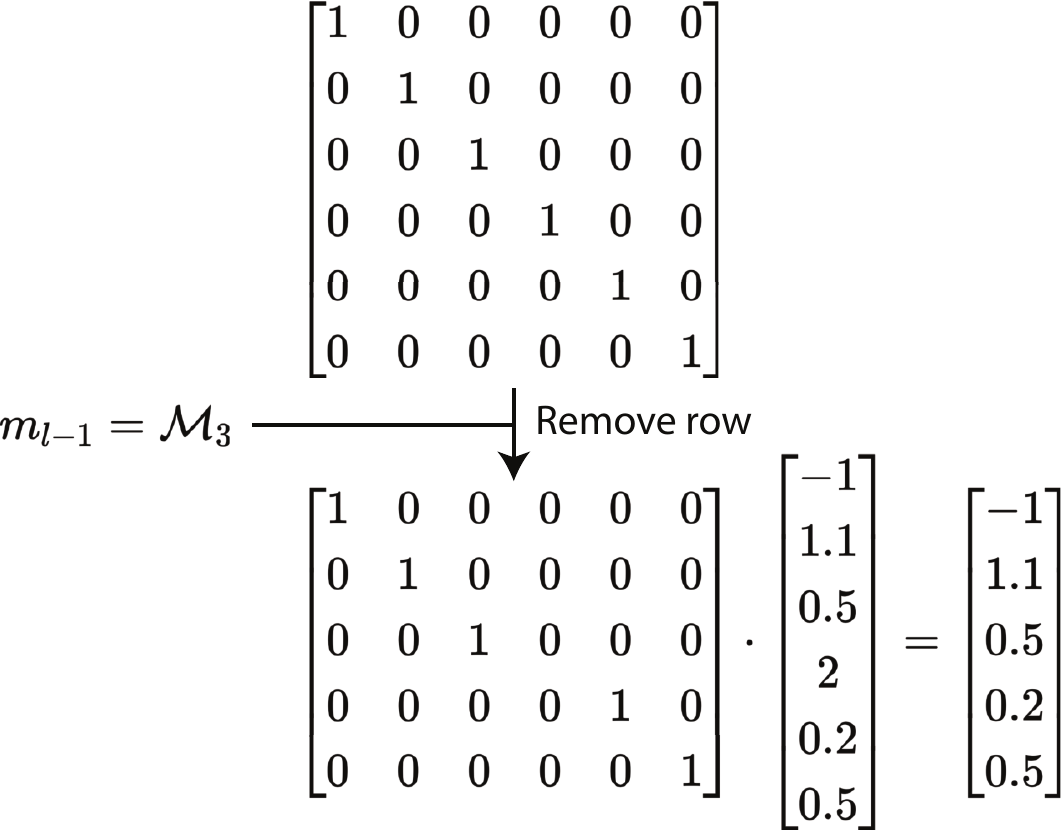}
        \caption{}
        \label{fig:gating}
    \end{subfigure}%
    \caption{Neural network architectures for generating optical multi-layer films. (a) We show one generation step in the plot. The hidden state $h_l$ of the GRU is passed to two MLPs to output material and thickness probabilities, respectively. The actual material and thickness for layer $l$ are sampled from categorical distributions parametrized by $p_{m_l}$ and $p_{d_l}$. Built-upon the baseline architecture, our proposed model adds a non-repetitive gating function and auto-regressive connection between the sampled material and the thickness MLP. (b) Illustration of how the non-repetitive gating works. Here we suppose there are a total of 5 materials. Thus, the gating matrix is of dimension $5\times 6$.}
\end{figure}

\subsubsection{Non-repetitive gating} 
The aforementioned material sampling procedure does not prevent the situation where the same material is sampled for adjacent layers. However, such consecutive layers of the same material are equivalent to a single thicker layer. Thus, allowing the sequence generator to generate the same material for adjacent layers leads to redundant computation. Moreover, doing so increases the exploration space size and makes the search problem harder. Thus, we introduce a non-repetitive gating function that removes the logit element corresponding to the most recently sampled material to prevent the sequence generator from generating the same materials in a row. This gating function is a matrix $I_{NR}\in \mathbb{R}^{|\mathcal{M}| \times (|\mathcal{M}|+1) }$ formed by removing the row corresponding to the most recently sampled material from an identify matrix. When multiplied with the logits vector $\sigma_{m_{l}}$, the element corresponding to that material will be removed, i.e., $\sigma_{m_l}' = I_{NR}\cdot \sigma_{m_l} \in \mathbb{R}^{|\mathcal{M}|}$. Then, we pass the transformed logit vector $\sigma_{m_l}'$ to the softmax layer to obtain the sampling probability. By doing so, we set the sampling probability for the recurring material to 0. With the non-repetitive gating, the generated material sequence is guaranteed to have different materials for adjacent layers. Note that, we do not apply the gating function for the first generation step because there is no previously sampled material.


\subsubsection{Auto-regressive generation of material and thickness} 
Because the proper thickness of a layer should depend on the material, we input the sampled material $m_l$ to the thickness MLP in addition to the hidden state $h_l$. A similar approach has been applied in RL problems where the actions are dependent on each other \cite{vinyals2019grandmaster}. Instead of using a one-hot vector to represent the material, we train a material embedding matrix $emb\in\mathbb{R}^{|\mathcal{M}|\times d}$ together with the sequence generator network. Each row $emb_m\in\mathbb{R}^d$ of the embedding matrix is a continuous representation of one material, where $d$ is the embedding size. Using an embedding allows us to use a large number of materials without significantly increasing the dimensionality of the material representation. The material embedding vector for the sampled material $emb_{m_l}$ is concatenated with the hidden state $h_{l}$ to form the input $[emb_{m_l}, h_{l}]$ to the material MLP.    

The full sequence generator architecture is plotted in Figure. \ref{fig:architecure}. To understand the effect of non-repetitive gating and modeling the dependency between the material and the thickness, we compare the proposed OML-PPO architecture against a baseline architecture Experiment section.

\subsection{Reinforcement learning training}

We train the sequence generation network with reinforcement learning. The goal of reinforcement learning is to maximize expected cumulative rewards $G = \mathbb{E}[\sum_{t=0}^\infty \gamma^t r_t]$ by learning a policy $\pi_\theta(a|s)$ that can map a state $s$ to an action $a$. Here, $\gamma$ is the discount factor that penalizes future rewards and $r_t$ is the reward at step $t$. The sequence generation network described above serves as the policy.

We represent the state at the $l$-th generation step as the concatenation of the last layer information and the GRU hidden state, i.e., $s_l = [(m_{l-1}, d_{l-1}), h_l]$. The actions $a_l$ correspond to the material and thickness $(m_{l}, d_{l})$ of the current layer. We set the reward to be 0 for all generation steps except the final step. At the final step (i.e., the structure $\mathcal{S}$ has been completely generated), we compute the spectrum of the generated structure with an optical spectrum calculation package TMM \cite{byrnes2016multilayer} and assign the final reward based on how well the structure spectrum matches with the target spectrum. We also tried to calculate the spectrum following every generation step and assign intermediate rewards. However, this dense-reward approach is slow and does not lead to improved performance. Thus, we only report the final-only approach here. We set the discount factor $\gamma =1$. Thus, the cumulative reward $G$ for the generated sequence $\mathcal{S}$ is simply the reward at the final step, which is defined as one minus the mean absolute error between the spectrum of the generated structure and the target spectrum:
\begin{equation}
    G(\mathcal{S}) = 1 - \frac{1}{K}\sum_{k=0}\frac{1}{J}\sum_{j=0}^{J-1}|T^{\mathcal{S}}(\lambda_j, \delta_k) - \tilde{T}(\lambda_j, \delta_k)|
    \label{eqn:reward}
\end{equation}
where $T^{\mathcal{S}}(\lambda_j, \delta_k)$ is the spectrum of the generated structure $\mathcal{S}$ at wavelength $\lambda_j$ under incidence angle $\delta_k$. Because $T \in [0, 1]$, the cumulative reward is always non-negative. The reward value will become higher as the spectrum $T^{\mathcal{S}}$ gets closer to the target spectrum $\tilde{T}$ until it reaches 1 when the structure spectrum perfectly matches with the target spectrum. 

During training, the sequence generator $\pi_\theta$ actively generates new structures and receive rewards. Our goal is to maximize the expected rewards for structures sampled from the sequence generation network:
\begin{equation}
    J(\theta)=\mathbb{E}_{\mathcal{S} \sim \pi_\theta}[G(\mathcal{S})].
\end{equation}

Based on the calculated rewards for generated sequences, the agent adjusts its parameters $\theta$ with gradient ascent so that future rewards can be improved. Here, we use a policy gradient algorithm to compute the gradient $\nabla_\theta J(\theta)$ for updating the sequence generator $\pi_\theta$. From the policy gradient theorem \cite{sutton2018reinforcement, schulman2017proximal}, we have
\begin{equation}
    g = \nabla_\theta J(\theta) = \mathbb{E}_{\mathcal{S}\sim \pi_\theta}\left[A(\mathcal{S}) \nabla_\theta \log P_\theta(\mathcal{S}) \right],
    \label{eqn:pg}
\end{equation}
where $P_\theta(\mathcal{S})=\prod_{l=0}^{N-1} p_\theta(m_l|s_{l-1}, h_{l-1})\cdot  p_\theta(d_l|m_l, s_{l-1}, h_{l-1})$ is the probability of sampling a structure $\mathcal{S}$ from the generator network $\pi_\theta$ and $A(\mathcal{S})$ is the estimated advantage function \cite{schulman2015high}, which measures the performance of the generated sequence $\mathcal{S}$ compared against the average performance of structures sampled from $\pi_\theta$.

\begin{figure}[ht]
    \centering
    \includegraphics[trim={0cm 1cm 0cm 0cm},clip,height=2in]{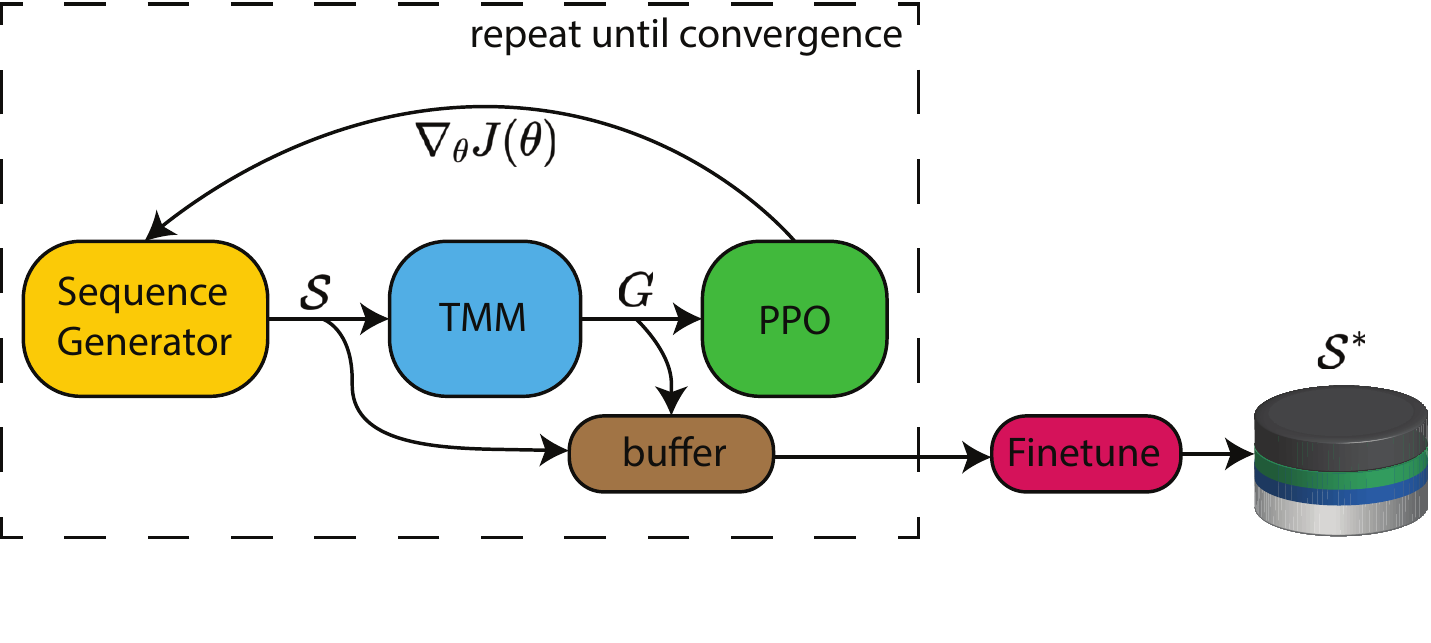}
    \caption{Pipeline of the sequence generator training process. We first generate multi-layer structures using the sequence generator $\pi_\theta$. The spectrum of the generated structures are simulated by the TMM module. Next, PPO algorithm is applied to compute the policy gradient $g$ for updating the sequence generator model. We keep pushing the best discovered structure into a buffer with size 1. This process is repeated until convergence. Finally, we finetune the layer thicknesses to obtain the design.}
    \label{fig:pipeline}
\end{figure}

Instead of directly updating the sequence generator using Eqn.\ref{eqn:pg}, we use a state-of-the-art policy gradient algorithm \textit{Proximal Policy Optimization} (PPO) \cite{schulman2017proximal} to compute the policy gradient from a surrogate objective function:
\begin{equation}
    g = \nabla_\theta \mathbb{E}_{\mathcal{S} \sim \pi_\theta}\left[\min \left(r(\theta) A_{\theta_v}(\mathcal{S}), \operatorname{clip}\left(r(\theta), 1-\epsilon, 1+\epsilon\right) A_{\theta_v}(\mathcal{S})\right)\right],
\end{equation}
where $r(\theta) = \frac{P_\theta(\mathcal{S})}{P_{\theta_\text{old}}(\mathcal{S})}$ is the importance weight that measures the distance between the policies before and after the gradient update. The $\operatorname{clip}$ function disincentivizes large update steps to the policy, where $\epsilon$ is a hyperparameter that affects the actual update size. Here, the advantage $A_{\theta_v}$ is estimated by \textit{Generalized Advantage Estimation} (GAE) \cite{schulman2015high}, which achieves a good balance between bias and variance of the estimated gradients. $\theta_v$ is the model parameters for a critic network that is trained together with the sequence generator. Compared to the vanilla policy gradient and actor-critic algorithms, PPO is more sample-efficient because it allows multi-step updates using the same batch of trajectories. Previous results show that PPO can achieve state-of-the-art performance on many tasks \cite{schulman2017proximal}. With the computed policy gradient, the sequence generator model parameters are updated using the Adam optimizer \cite{kingma2014adam}. The model training process is summarized in Figure. \ref{fig:pipeline}. Similar to the \textit{active search} approach in Bello et al. \cite{bello2016neural}, we output the best structure discovered throughout the entire training process as the final design.  The pseudocode that summarizes our design generation process is given in Algorithm \ref{alg:omlppo}.

Our model is implemented using PyTorch \cite{paszke2019pytorch} and Spinning Up \cite{SpinningUp2018}. The data used in this study and our code are publicly available\footnote[1]{\href{https://github.com/hammer-wang/oml-ppo}{https://github.com/hammer-wang/oml-ppo}}.


\begin{algorithm}[ht]
\SetAlgoLined
\KwIn{target $\tilde{T}$, number of epochs $K$, batch size $B$, maximum length $L$}
\KwOut{Optical multi-layer sequence $\mathcal{S}^*$}
    Initialize sequence generator parameters $\theta$\;
    
    Initialize critic network parameters $\theta_v$\;
    
    Initialize best design $\mathcal{S}^*$\;
 
     \For{k = 1, \ldots, K}{
        $\mathcal{S}_i \sim \texttt{SampleDesign}(L, B, \theta)$ \;
        
        $\mathcal{S}^* \leftarrow \texttt{SelectBest}(\{\mathcal{S}_i\}, \mathcal{S}^*, \tilde{T})$
        
        $\theta, \theta_v \leftarrow \texttt{PPOUpdate}(\{\mathcal{S}_i\}, \theta, \theta_v)$\;
        }
    
    $\mathcal{S}^* \leftarrow \texttt{QuasiNewton}(\mathcal{S}^*, \tilde{T})$  
\caption{OML-PPO}
\label{alg:omlppo}
\end{algorithm}

\section{Experiment}
We applied the proposed method to two optical design tasks that are relevant to energy applications, i.e., 1) designing ultra-wideband absorbers and 2) designing incandescent light bulb filters. The designed ultra-wideband absorbers can help solar thermal panels to absorb the sunlight more efficiently and the light bulb filter can enhance incandescent light bulb efficiency in emitting visible light while suppressing the radiation in the infrared range that represents energy loss. We also did an ablation study to understand the effect of non-repetitive gating and auto-regressive materials/thickness sampling. 

\textbf{Performance evaluation}: In task 1 ultra-wideband absorber design, we measure the quality of the designed structure by \textit{average absorption}. In task 2 incandescent light bulb filter, we calculate the visible light \textit{enhancement factor} to measure the performance of designed structures.

\subsection{Task 1: ultra-wideband absorber}
Firstly, we apply our algorithm to the task of designing an ultra-wideband absorber for the wavelength range [400, 2000] nm. We choose the target spectrum as a constant 100\% absorption under normal light incidence angle (i.e., the light is shining at the absorber at a right angle) to represent an ideal broadband absorber. This task has been previously studied by Yang et al. \cite{yang2016compact} based on physical models, where the broadband absorption is achieved by overlapping multiple absorption resonances and with an overall graded-index structure to minimize reflection. The authors designed a 5-layer structure using MgF\textsubscript{2}, TiO\textsubscript{2}, Si, Ge, and Cr. The simulated average absorption of their structure over the wavelength range is 95.37\% under normal incidence. If not specified otherwise, we assume normal incidence when reporting average absorption. 

\begin{table}[h]
    \centering
    \caption{Available materials for constructing the ultra-wideband absorber.}
    \begin{tabular}{c c c c c c c c}
        \hline
        Ag & Al& Al\textsubscript{2}O\textsubscript{3}& Cr& Fe\textsubscript{2}O\textsubscript{3} & Ge& HfO\textsubscript{2} & MgF\textsubscript{2} \\
        Ni & Si& SiO\textsubscript{2}& Ti & TiO\textsubscript{2} & ZnO & ZnS & ZnSe \\
        \hline
    \end{tabular}
    \label{tab:materials}
\end{table}

We hypothesize that, when choosing from a larger set of materials than used in the previous work \cite{yang2016compact}, it is possible to design a structure with higher average absorption than the human-designed structure. Thus, we expanded the original material set \cite{yang2016compact} to include 11 more materials (16 total). The set of materials is listed in Table \ref{tab:materials}. We set the available discrete thicknesses $\mathcal{D}$ to be $\{15, 20, 25, \ldots, 200\}$ nm with a total of 38 different values. When training the sequence generator, we set the learning rate to $5\times 10^{-5}$ and the maximum length to $L=6$. The material embedding size $d$ is set to 5, i.e., $emb_m \in \mathbb{R}^5$. The generator is trained for a total of $3,000$ epochs with the batch size set to be $1,000$ generation steps. We repeat the training for $10$ runs with different random seeds. The best structure discovered in each run was recorded and finetuned using the quasi-Newton method.

\begin{table}[ht]
    \centering
    \caption{RL designed 14-layer structure with 99.24\% average absorption.}
    \begin{tabular}{|l l l|l l l|}
    \hline
    \textbf{ID} & \textbf{Material} & \textbf{Thickness} &  \textbf{ID} & \textbf{Material} & \textbf{Thickness}\\
    1 & \mgf & 123 nm & 8  & Si   & 15 nm\\
    2 & \tio & 32 nm & 9  & Cr   & 17 nm\\
    3 & \mgf & 21 nm & 10 & Ge   & 15 nm\\
    4 & Si   & 15 nm & 11 & \tio & 33 nm\\
    5 & \tio & 15 nm & 12 & Cr   & 29 nm\\ 
    6 & Si   & 15 nm & 13 & \tio & 81 nm\\
    7 & Ge   & 15 nm & 14 & Cr   & 116 nm\\
    \hline
    \end{tabular}
    \label{tab:14layer}
\end{table}

\begin{figure}[ht]
    \centering
    \begin{subfigure}[t]{0.45\textwidth}
        \centering
        \includegraphics[height=1.7in]{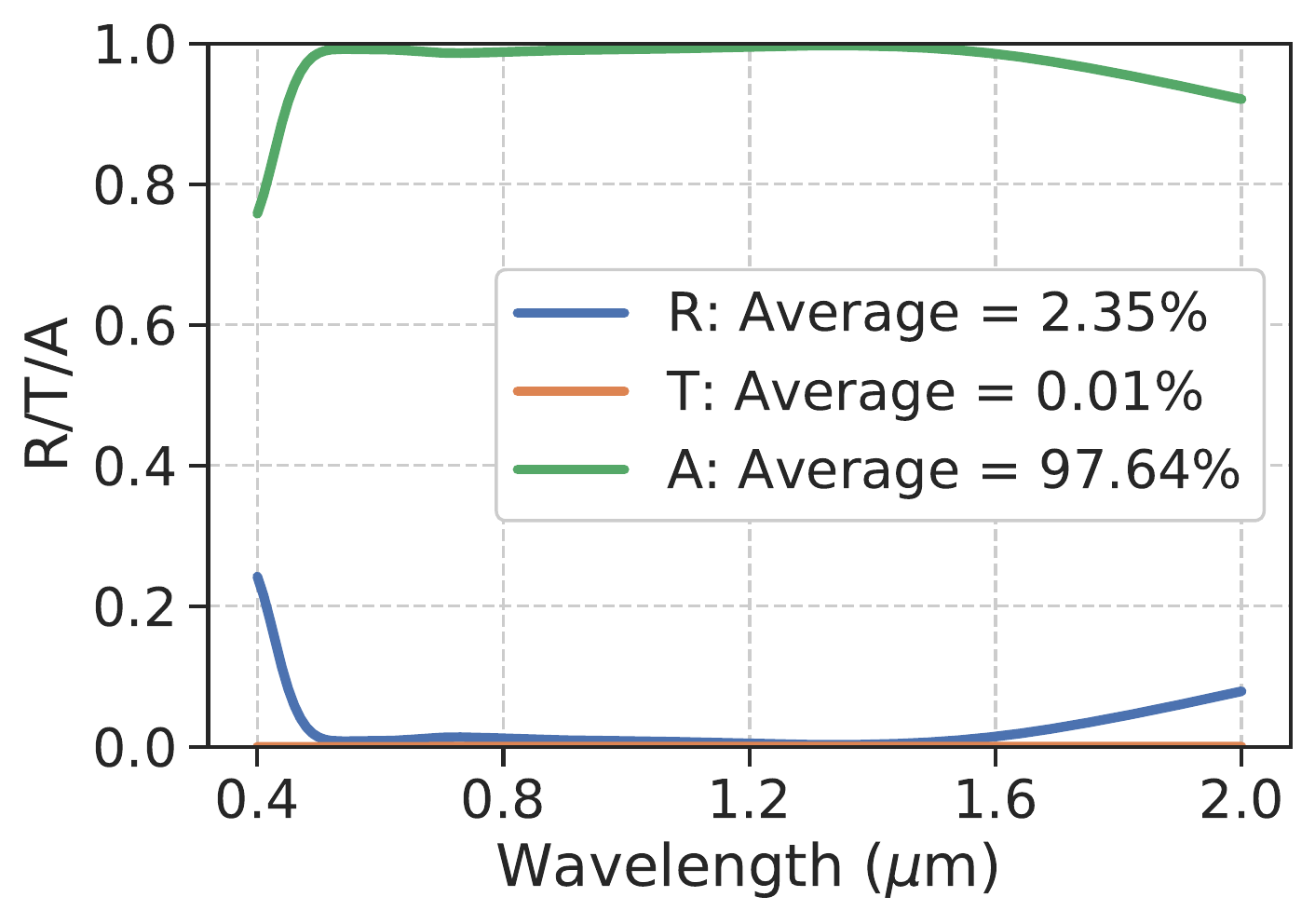}
        \caption{}
        \label{fig:maxlen6_normal}
    \end{subfigure}%
    ~ 
    ~
    \begin{subfigure}[t]{0.45\textwidth}
        \centering
        \includegraphics[height=1.7in]{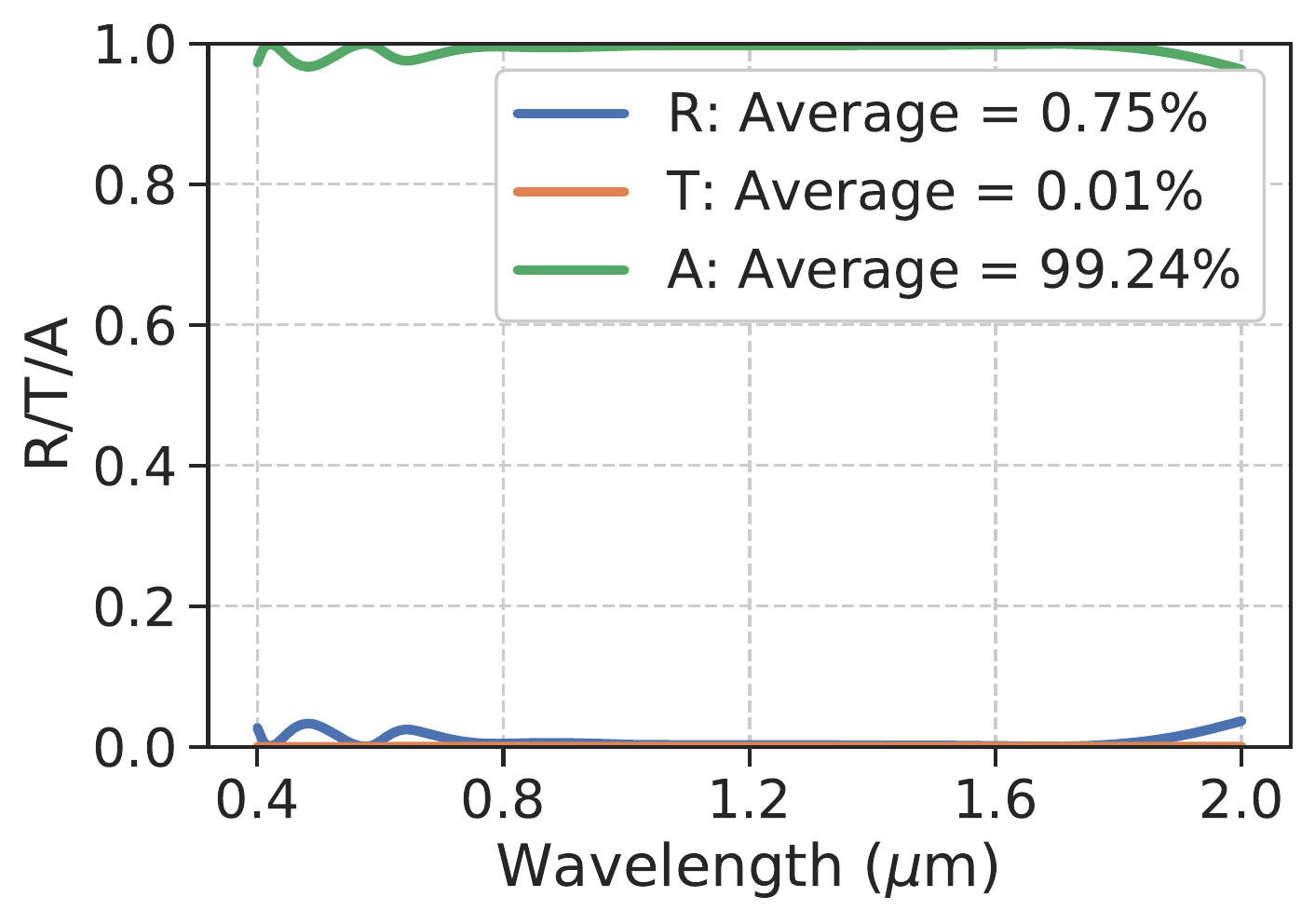}
        \caption{}
        \label{fig:maxlen14_normal}
    \end{subfigure}%
    ~ 
    
    \caption{Normal incidence spectrum for the best discovered absorber structures with 5 and 14 layers. R: reflection, T: transimission, A: absorption. We design the multi-layer thin film to have high \textit{absorption} over the entire wavelength range. (a) Normal incidence spectrum for the 5-layer structure. (b) Normal incidence spectrum for the 14-layer structure.}
    \label{fig:absorber_res}
\end{figure}

It is worth noting that our algorithm can yield very similar structures as that reported in  \cite{yang2016compact}, i.e., it can search for and find the structure designed based by human experts. One of such structures is \{(MgF2, 112 nm), (TiO2, 55 nm), (Ti, 30 nm), (Ge, 30 nm), (Cr, 200 nm)\} with an average absorption of 96.12\%, which has exactly the same material composition as the one reported previously \cite{yang2016compact}. However, the best structure discovered by the algorithm, exhibiting a higher average absorption of 97.64\%, is \{(SiO2, 115 nm), (Fe2O3, 70 nm), (Ti, 15 nm), (MgF2, 124 nm), (Ti, 148 nm)\}. The spectrum under normal incidence are plotted in Figure \ref{fig:maxlen6_normal}.

\begin{figure}[h]
    \centering
    \begin{subfigure}[t]{0.45\textwidth}
        \centering
        \includegraphics[height=1.7in]{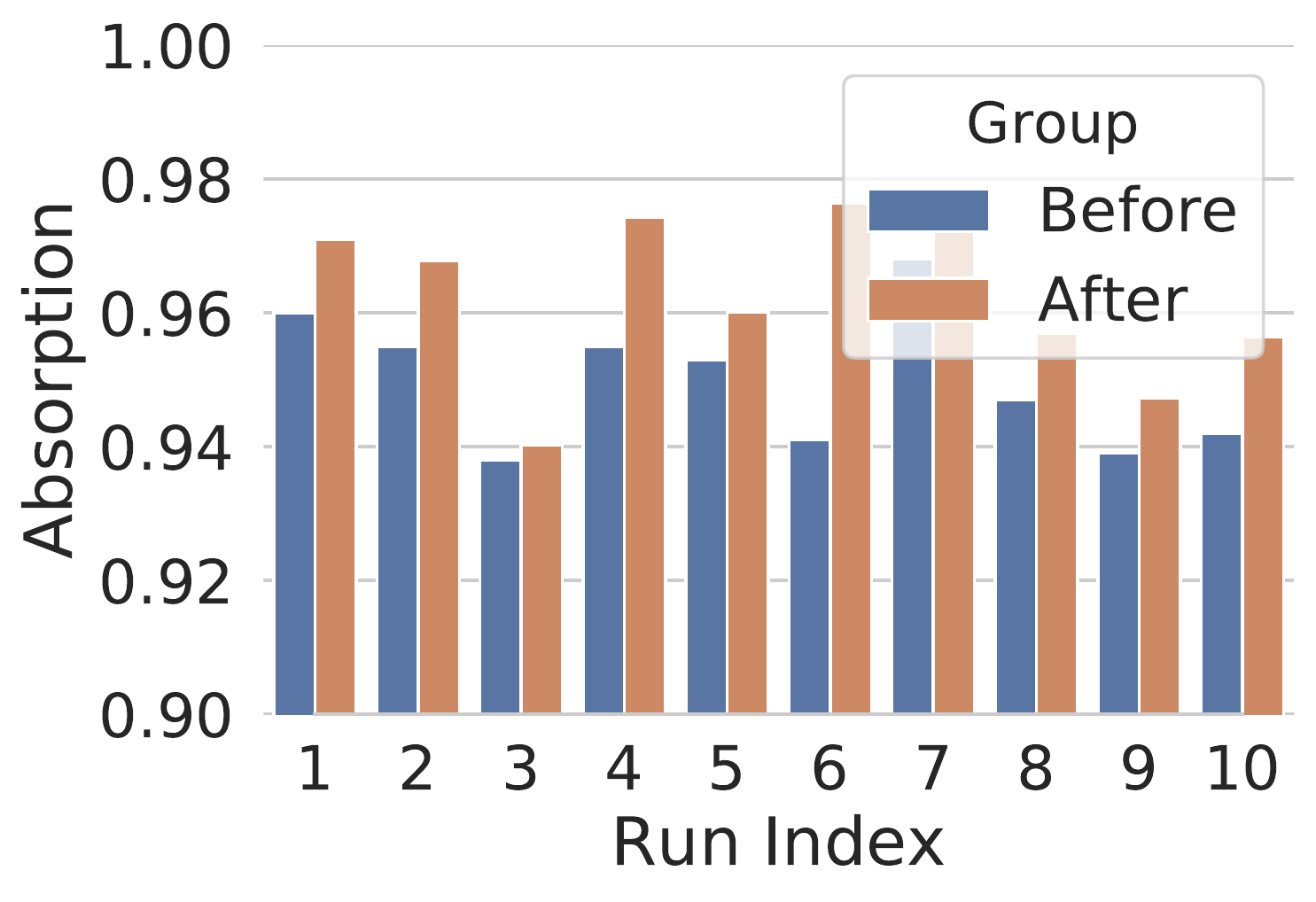}
        \caption{}
    \end{subfigure}%
    ~ 
    \begin{subfigure}[t]{0.45\textwidth}
        \centering
        \includegraphics[height=1.7in]{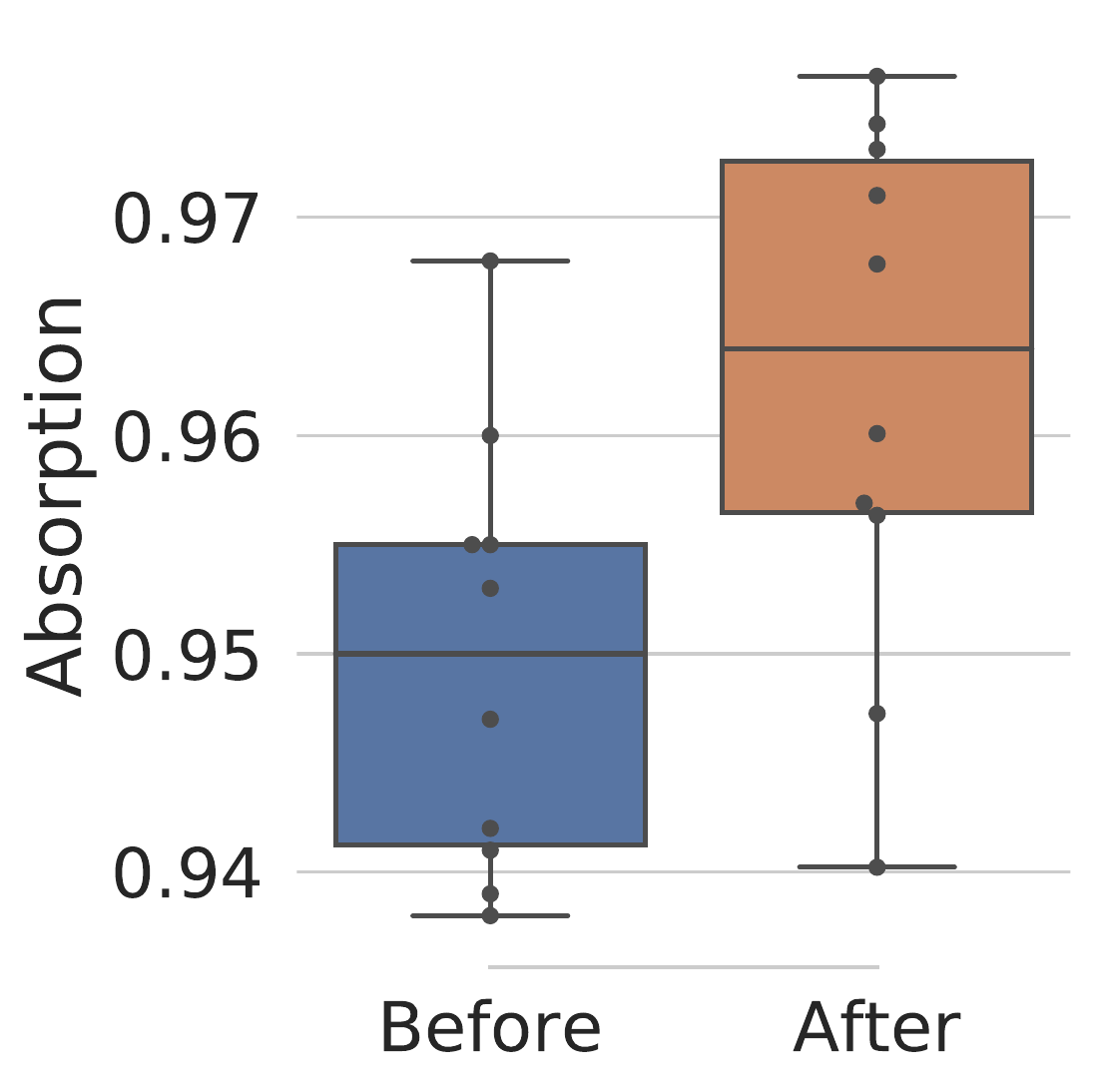}
        \caption{}
    \end{subfigure}%
    
    \caption{Absorption values before and after finetuning. finetuning improves the average absorption of every structure discovered in each run. (a) Average absorption values before and after finetuning for each individual run. (b) Box-plot for ten average absorptions values}
    \label{fig:finetune}
\end{figure}

We plot the best absorption values before and after finetuning of all ten runs in Figure. \ref{fig:finetune}. After finetuning, the average absorptions for the discovered structures across all runs were improved. We found that the algorithm is robust to the randomness during training as 8 out of the 10 runs achieved an absorption that is higher than 95\% after finetuning. 

In an additional experiment, we explore whether the algorithm can design a structure with more layers to achieve even higher absorptions. We set the maximum length $L=15$ and sample layer materials from MgF\textsubscript{2}, TiO\textsubscript{2}, Si, Ge, and Cr. The best discovered structure has 14 layers with an average absorption of 99.24\%. The structure configuration is summarized in Table \ref{tab:14layer}. We plot the normal incidence spectrum structure in Figure. \ref{fig:maxlen14_normal}. The structure discovered by OML-PPO reaches close-to-perfect performance under normal incidence and has high absorption over a wide range of angles.


\subsection{Task 2: incandescent light bulb filter}
\begin{figure*}[ht]
    \centering
    \begin{subfigure}[t]{0.45\textwidth}
        \centering
        \includegraphics[height=1.7in]{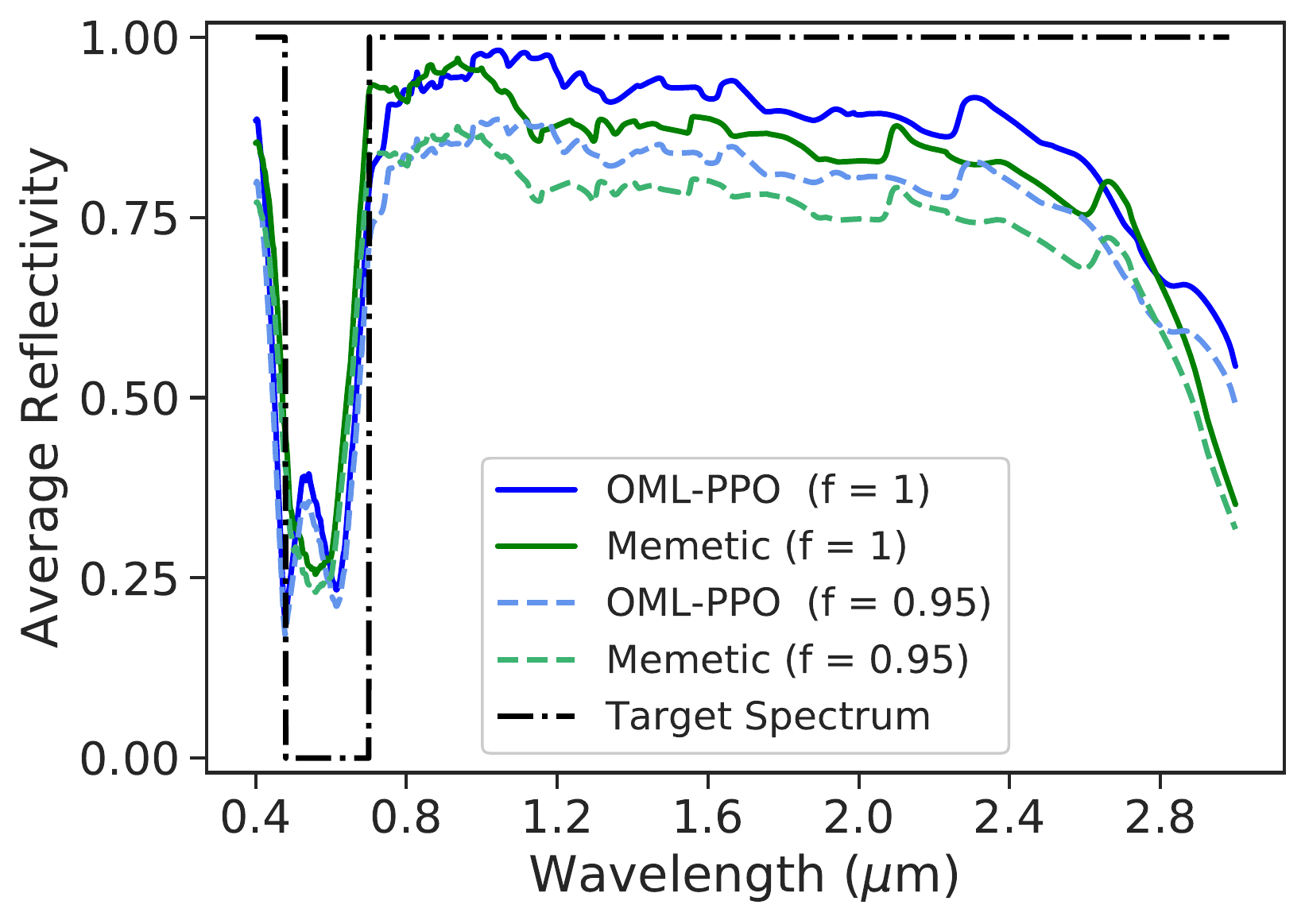}
        \caption{}
        \label{fig:incandescent_spectrum}
    \end{subfigure}
    ~ 
    \begin{subfigure}[t]{0.45\textwidth}
        \centering
        \includegraphics[height=1.9in]{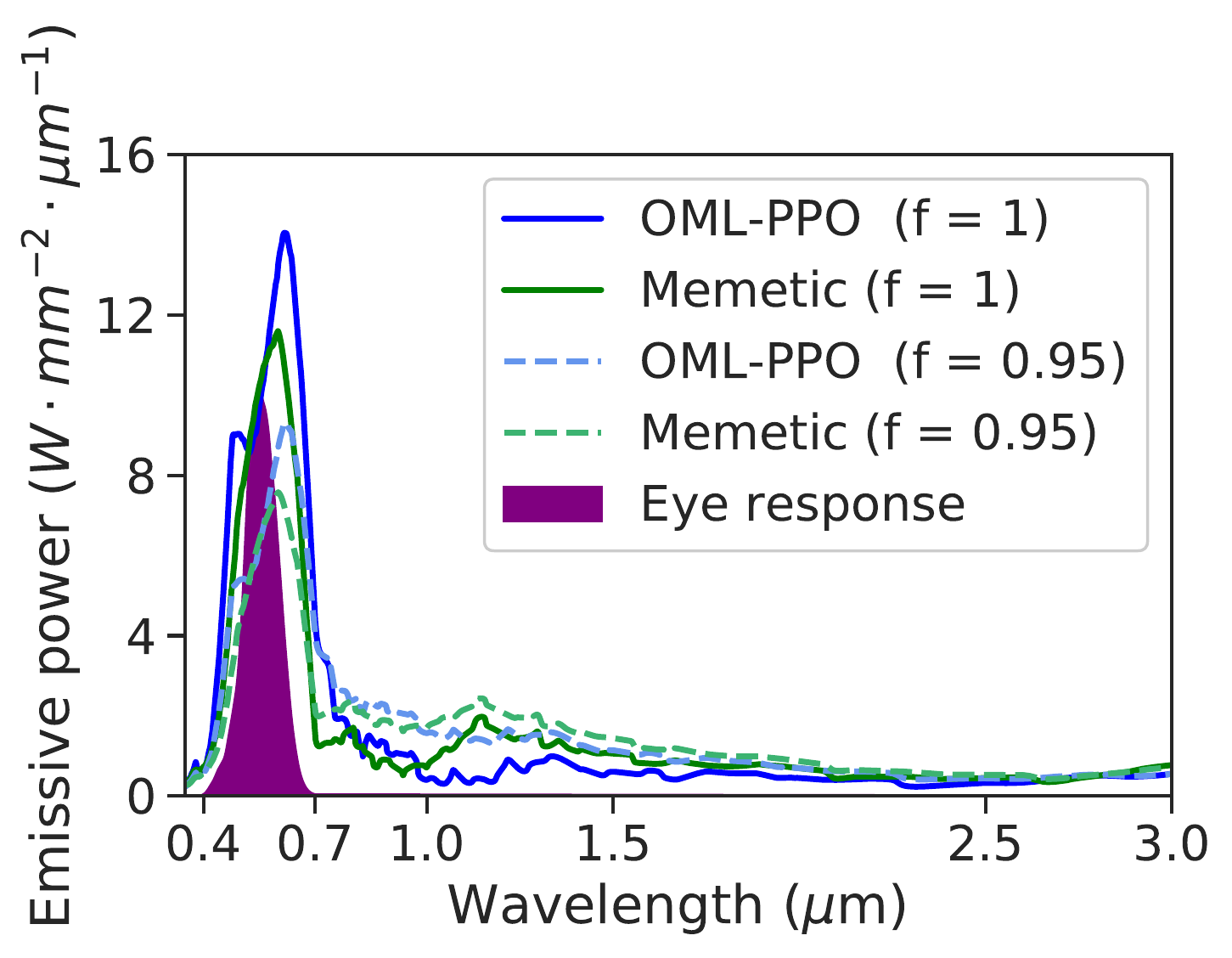}
        \caption{}
    \end{subfigure}
    
    \caption{Results on the incandescent light bulb design. (a) Target spectrum and the average reflectivity of structures designed by OML-PPO and the memetic algorithm. (b) Emissive power spectrum. A good design will have high emissive power in the visible range [380, 780] nm. $f$ is the view factor that equals the proportion of emitted light from the light bulb filament that can reach the light bulb filter. We report results under view factors 0.95 and 1.}
    \label{fig:incandescent_res}
\end{figure*}

To further test whether our method is scalable to more complicated tasks, we apply the proposed method for designing a filter that can enhance the luminous efficiency of incandescent light bulbs \cite{zhou2016efficient, ilic2016tailoring}. The idea is to reflect the infrared light emitted by the light bulb filament so that its energy can be recycled. To this end, we set the target reflectivity to be 0\% in the range [480, 700] nm, and 100\% outside this range (Figure. \ref{fig:incandescent_spectrum}). In this way, the infrared light, which cannot contribute to lighting, will be reflected back to heat up the emitter. 

A similar design has been previously studied \cite{ilic2016tailoring, shi2017optimization}. We choose the same seven dielectric materials as the available materials: Al\textsubscript{2}O\textsubscript{3}, HfO\textsubscript{2}, MgF\textsubscript{2}, SiC, SiO\textsubscript{2}, and TiO\textsubscript{2} \cite{shi2017optimization}. Similar to our previous experiment, we train our policy for $10$ runs with different random seeds. Here, we set the maximum allowed length $L = 45$ and the learning rate to be $5 \times 10^{-5}$. The number of epochs and batch size are 10,000 and 3,000, respectively. The best discovered structure is reported in Appendix.

In Figure \ref{fig:incandescent_res}, we compare the average reflectivity normalized over all incidence angles (0 - 90 degree) of the 42-layer structure designed with our algorithm and the 41-layer structure designed by a memetic algorithm \cite{shi2017optimization}. Our structure has a higher average reflectivity in the infrared range ($>780$ nm) than the 41-layer structure. 

We quantitatively evaluated the performance of the designed filter by calculating the enhancement factor for visible light (400 - 780 nm) under a fixed operating power. The results are reported in Table \ref{tab:enhancement}. Details about the calculation of enhanced factor is included in Appendix. 

\begin{table}[ht]
    \centering
    \caption{Visible light enhancement. Our RL-designed structure achieved 8.5\% higher visible light enhancement than the structure designed by a memetic algorithm.}
    \begin{tabular}{|c|c|}
    \hline
    \textbf{Model} & \textbf{Enhancement factor} \\
    \hline
    OML-PPO & $\mathbf{16.60}$\\
    \hline
    Memetic \cite{shi2017optimization} & $15.30$\\
    \hline
    \end{tabular}
    \label{tab:enhancement}
\end{table}

\subsection{Ablation study} On the ultra-wideband absorber design task, we conducted an ablation study to understand the effect of non-repetitive gating and auto-regressive generation of materials and thicknesses. We trained four different models: 1) OML-PPO with both non-repetitive gating and auto-regressive generation, 2) non-repetitive gating only, 3) auto-regressive generation only, 4) neither non-repetitive gating nor the auto-regressive generation. For each model, we repeated the training for ten times. The maximum absorption values discovered by each model before finetuning are reported in Table \ref{tab:compare}. Both non-repetitive gating and the auto-regressive material/thickness generation improve the performance of the baseline model. 

\begin{figure}[ht]
    \centering
    \begin{subfigure}[t]{0.4\textwidth}
        \centering
        \includegraphics[height=1.5in]{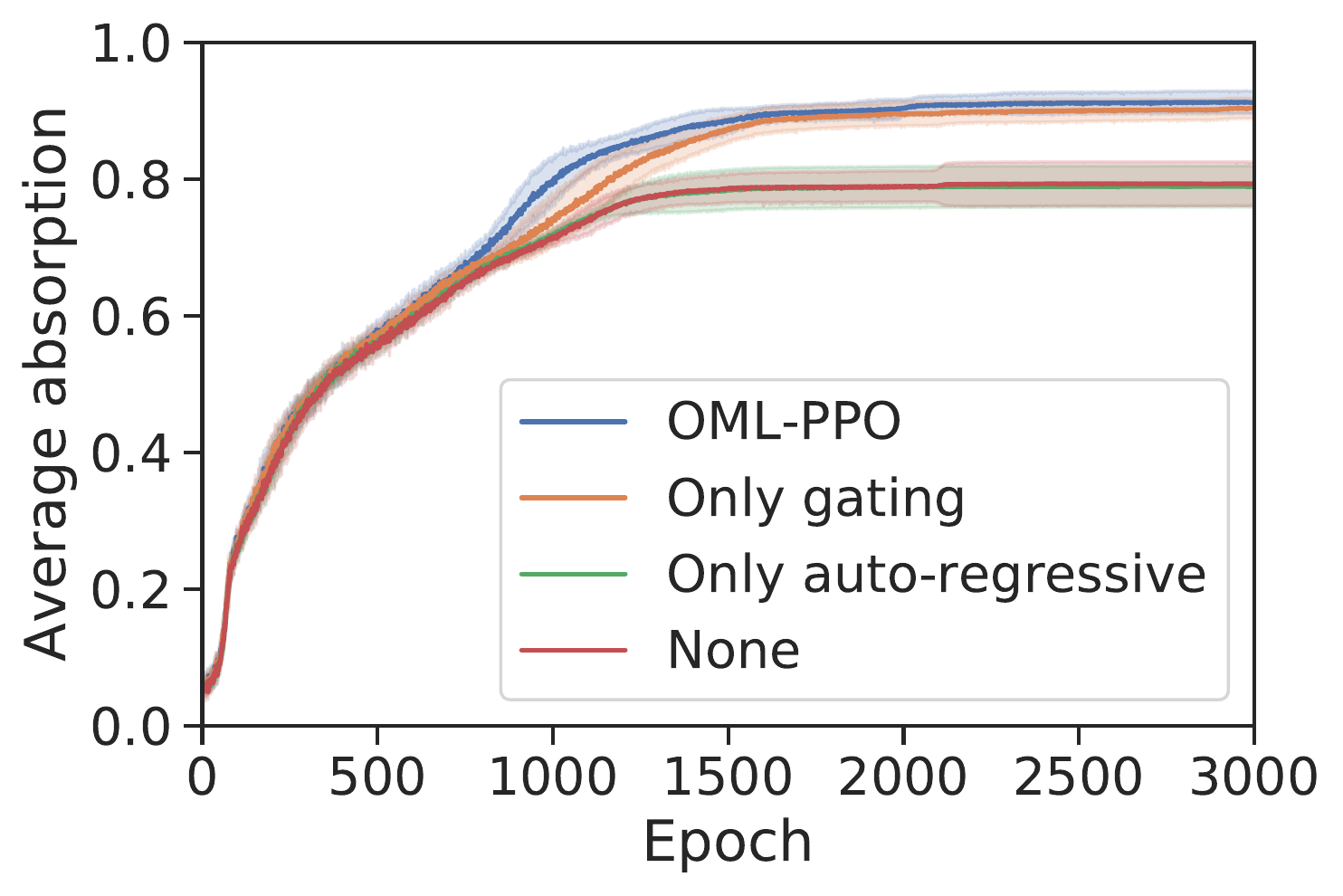}
        \caption{}
    \end{subfigure}%
    ~
    \begin{subfigure}[t]{0.4\textwidth}
        \centering
        \includegraphics[height=1.5in]{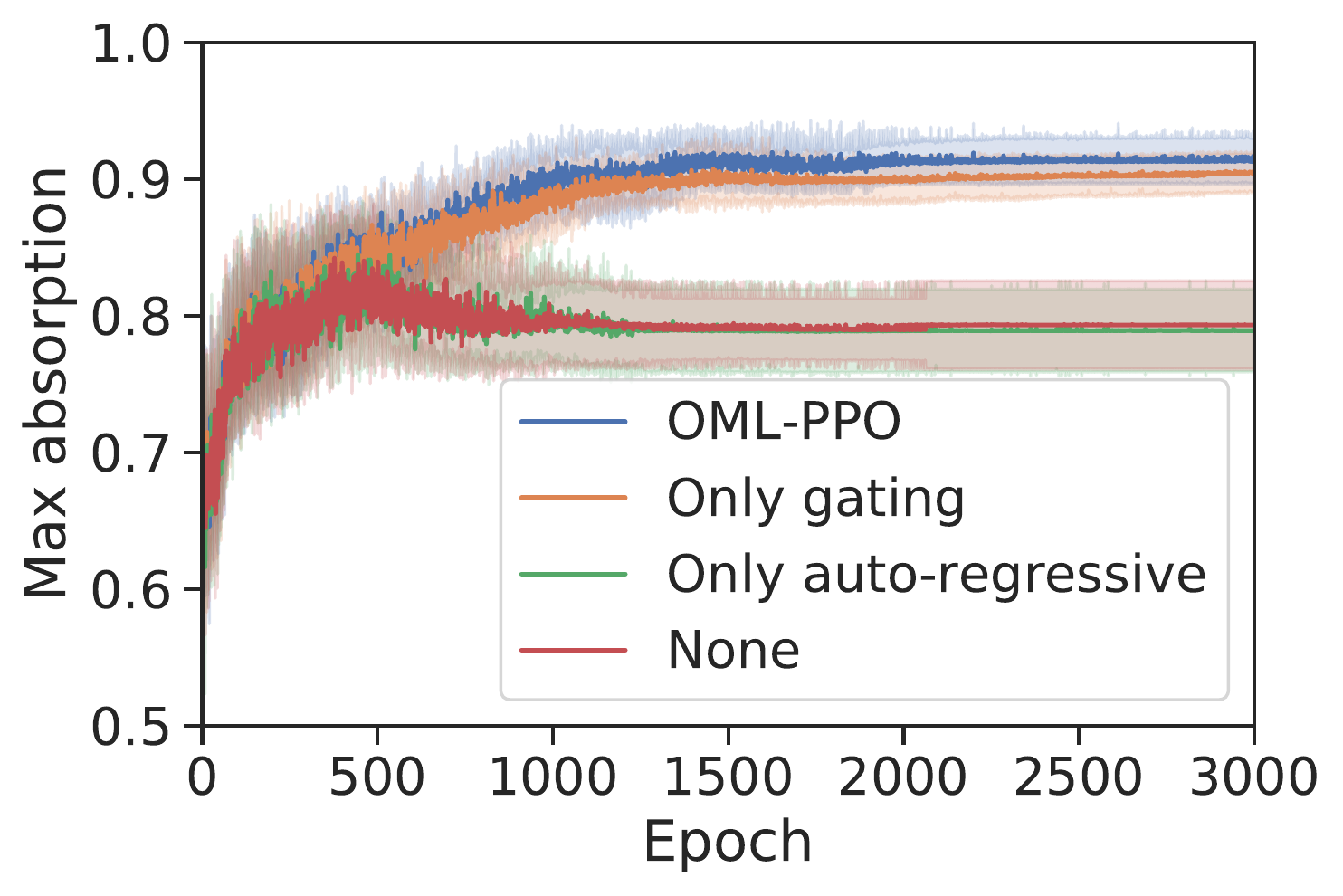}
        \caption{}
    \end{subfigure}%
    
    \caption{Training trajectory of OML-PPO and other baseline algorithms. (a) Average absorption trajectory. (b) Maximum absorption trajectory. The non-repetitive gating enables the model to converge to better solutions than models without the gating. The shaded area corresponds to one standard deviation.}
    \label{fig:training_traj}
\end{figure}

\begin{table}[ht]
    \centering
    \caption{Highest absorption values discovered by each algorithm across 10 runs. The mean average absorption values and standard deviations of the 10 runs are reported.}
    \begin{tabular}{|c|c|}
    \hline
    \textbf{Model} & \textbf{Average Absorption} \\
    \hline
    OML-PPO & $\mathbf{94.98\%\pm 0.99\%}$\\
    \hline
    Only gating & $94.05\%\pm 1.39\%$\\
    \hline
    Only auto-regressive & $91.55\%\pm 1.14\%$\\
    \hline
    None (baseline) & $91.03\%\pm 0.87\%$ \\
    \hline
    \end{tabular}
    \label{tab:compare}
\end{table}

In Figure. \ref{fig:training_traj}, we plot the average absorption and maximum absorption of the structures generated in each epoch over the entire training trajectory. The effect of non-repetitive gating is more significant than auto-regressive material/thickness generation as the OML-PPO and the only-gating variants both significantly outperform the other two variants. The non-repetitive gating significantly improves the model convergence during training. When non-repetitive gating and the auto-regressive sampling are combined together, the model achieves the best performance. 

\section{Conclusion}
We introduced a novel sequence generation architecture and a deep reinforcement learning pipeline to automatically design optical multi-layer films. To the best of our knowledge, our work is the first to apply deep reinforcement learning to design multi-layer optical structures with the optimal number of layers not known beforehand. Using a sequence generation network, the proposed method can select material and thickness for each layer of a multi-layer structure sequentially. On the task of designing an ultra-wideband absorber, we demonstrate that our method can achieve high performance robustly. The algorithm automatically discovered a 5-layer structure with 97.64\% average absorption over the [400, 2000] nm range, which is 2\% higher than a structure previously designed by human experts. When applied to generate a structure with more layers, the algorithm discovered a 14-layer structure with 99.24\% average absorption, approaching perfect performance. On the task of designing incandescent light bulb filters, our method achieves 8.5\% higher visible light enhancement factor than a structure designed by a state-of-art memetic algorithm. 

Through an ablation study, we showed that customizing the sequence generation network based on optical design domain knowledge can greatly improve the optimization performance. Our results demonstrated the high performance of the proposed method on complicated optical design tasks. Because the proposed method does not rely on hand-crafted heuristics, we believe that it can be applied to many other multi-layer optical design tasks such as lens design and multi-layer metasurface design.  

\bibliographystyle{iclr2020_conference}
\bibliography{iclr2020_conference.bib}

\newpage
\appendix

\section{RL-designed 42-layer incandescent light bulb}
\label{appendix:42layer}
{\setlength{\tabcolsep}{5pt}
\begin{table*}[ht!]
    \centering
    \caption{RL designed incandescent light bulb filter with 42 layers. The total thickness is  \SI{8.54}{\micro\metre}.}
    \small
    \begin{tabular}{|l l l|l l l|l l l|}
    \hline
    \textbf{ID} & \textbf{Material} & \textbf{Thickness} &  \textbf{ID} & \textbf{Material} & \textbf{Thickness} & \textbf{ID} & \textbf{Material} & \textbf{Thickness} \\
    1  & \sio & 289 nm & 15 & SiC  & 210 nm & 29 & SiC  & 117 nm\\
    2  & SiN  & 268 nm & 16 & SiN  & 168 nm & 30 & \mgf & 224 nm \\
    3  & \mgf & 185 nm & 17 & \mgf & 200 nm & 31 & SiC  & 122 nm\\
    4  & SiN  & 189 nm & 18 & SiC  & 227 nm & 32 & \mgf & 235 nm\\
    5  & SiC  & 214 nm & 19 & SiN  & 242 nm & 33 & SiC  & 127 nm\\ 
    6  & SiN  & 214 nm & 20 & \mgf & 222 nm & 34 & \mgf & 230 nm\\
    7  & \mgf & 210 nm & 21 & SiC  & 228 nm & 35 & SiC  & 234 nm\\
    8  & SiN  & 206 nm & 22 & \mgf & 216 nm & 36 & \mgf & 218 nm\\
    9  & SiC  & 205 nm & 23 & SiC  & 229 nm & 37 & SiC  & 235 nm\\
    10 & SiN  & 183 nm & 24 & \mgf & 203 nm & 38 & \mgf & 220 nm\\
    11 & \mgf & 184 nm & 25 & SiC  & 101 nm & 39 & SiC  & 231 nm\\
    12 & SiN  & 179 nm & 26 & \mgf & 209 nm & 40 & \mgf & 216 nm\\
    13 & SiC  & 203 nm & 27 & SiC  & 121 nm & 41 & SiC  & 233 nm\\
    14 & SiN  & 273 nm & 28 & \mgf & 225 nm & 42 & \alo & 95 nm\\
    \hline
    \end{tabular}
    \label{tab:42layer}
\end{table*}
}

\section{Visible light enhancement factor}
We first calculated the angle averaged emissivity $\epsilon_{\text{avg}}(\lambda)$ over a hemisphere:
\begin{align*}
    \epsilon_{\text{avg}}(\lambda) = \frac{2\pi\int_{0}^{\pi/2} \cos\delta\cdot \sin\delta\cdot \epsilon_{\text{eff}}(\lambda, \delta) d\delta}{2\pi\int_{0}^{\pi/2} \cos\delta\cdot \sin\delta d\delta} \\ = 2\int_{0}^{\pi/2} \cos\delta\cdot \sin\delta\cdot \epsilon_{\text{eff}}(\lambda, \delta) d\delta,
\end{align*}
where $\epsilon_\text{eff}(\lambda, \delta) = 1 - f^2 R(\lambda, \delta)$. $R(\lambda, \delta)$ is the reflection of the structure at wavelength $\lambda$ under the incidence angle of $\delta$.  $f$ is the view factor that equals to the proportion of the light from the emitter that can reach the filter. We compared two different view factors $f = 1$ and $0.95$ in our calculation. In addition, we assume the light bulb operates at 100 W and the surface area of the emitter is equal to $Area = 20\;\text{mm}^2$. Then, we can solve for the temperature $t$ of the light emitter with the equation:
\begin{equation*}
P_{\text {emitter }}(t)= Area \cdot \int I_{\text {emitter}}(\lambda, t) \epsilon_{\text {eff }}(\lambda) d\lambda, 
\end{equation*}
where $I_{\text {emitter}}(\lambda, t)=\frac{2 h c^{2}}{\lambda^{5}} \frac{1}{e^{h c /\left(\lambda k_{B} t\right)}-1}$ is the blackbody emission intensity spectrum. With view factor $f = 1 \; (0.95)$, the OML-PPO designed filter leads to the emitter temperature of 3810 K (3553 K) while the structure designed by the memetic algorithm achieves a temperature of 3750 K (3498 K). The black body temperature under the same condition is calculated to be $t_0 = 2578$ K. We measure the enhancement factor by:
\begin{equation*}
    \chi=\frac{\int \epsilon_{\mathrm{eff}}(\lambda) I_{\mathrm{emitter} }(\lambda, t) V(\lambda) \mathrm{d} \lambda}{\int I_{\mathrm{emitter}}(\lambda, t_0) V(\lambda) \mathrm{d} \lambda},
\end{equation*}
where $V(\lambda)$ is the human eye's sensitity spectrum \citep{sharpe2005luminous}. Our structure achieves an enhancement factor of 16.60 (10.67) while the memetic structure has an enhancement factor of 15.30 (9.72). The 42-layer structure designed by OML-PPO outperforms the previous 41-layer design by 8.5\% (9.8\%) in terms of the visible light enhancement.

\newpage
\section{Angle-dependent absorption map for ultra-wideband absorbers}
\begin{figure}[ht]
    \centering
    \begin{subfigure}[t]{0.4\textwidth}
        \centering
        \includegraphics[height=1.5in]{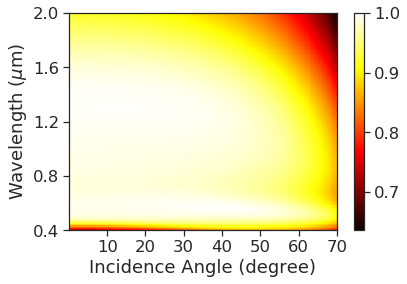}
        \caption{}
        \label{fig:maxlen6_angle}
    \end{subfigure}%
    ~
    \begin{subfigure}[t]{0.4\textwidth}
        \centering
        \includegraphics[height=1.5in]{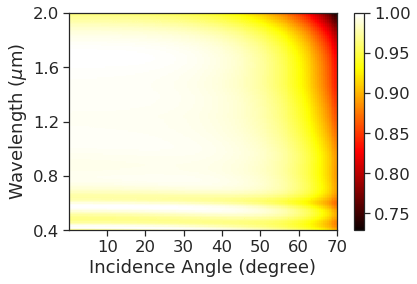}
        \caption{}
        \label{fig:maxlen14_angle}
    \end{subfigure}%
    
    \caption{Angle-dependent absorption map for the best discovered absorber structures with 5 and 14 layers. Both achieves high absorption over a wide range of angles. (a) 5-layer structure. (b) 14-layer structure.}
    \label{fig:absorber_appendix_task1}
\end{figure}

\section{Angle-dependent reflection map for incandescent light bulb filter}
\begin{figure}[ht]
    \centering
    \begin{subfigure}[t]{0.4\textwidth}
        \centering
        \includegraphics[height=1.5in]{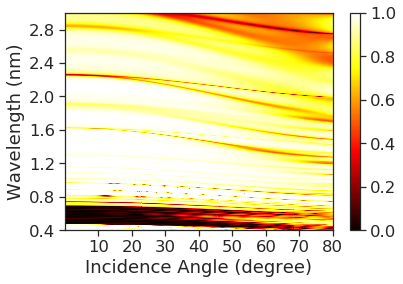}
        \caption{}
        \label{fig:angle_dependent_omlppo}
    \end{subfigure}
    ~
    \begin{subfigure}[t]{0.4\textwidth}
        \centering
        \includegraphics[height=1.5in]{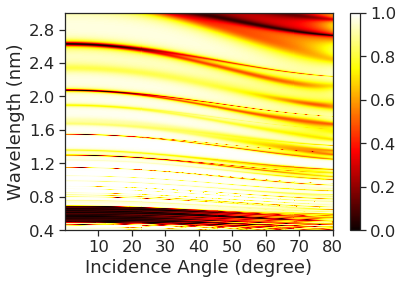}
        \caption{}
        \label{fig:angle_dependent_memetic}
    \end{subfigure}
    
    \caption{Angle-dependent reflection map for RL-designed structure and a structure designed by memetic algorithm. (a) structure designed by RL algorithm. (b) structure designed by memetic algorithm.}
    \label{fig:absorber_appendix_task2}
\end{figure}

\end{document}